Declarations of interest: none

# Tuning the structure of *in*-situ synthesized few layer graphene/carbon composites into nanoporous vertically aligned graphene electrodes with high volumetric capacitance.

Anurag Mohanty,[a] Izabela Janowska[a]*

[a] Institut de Chimie et Procédés pour l'Énergie, l'Environnement et la Santé (ICPEES), CNRS UMR 7515-University of Strasbourg, 25 rue Becquerel 67087 Strasbourg, France

*Corresponding author: tel: +33 3 68 85 26 33, e-mail: janowskai@unistra.fr

Tuning the structure of *in*-situ synthesized few layer graphene/carbon composites into nanoporous vertically aligned graphene electrodes with high volumetric capacitance.

Anurag Mohanty,[a] Izabela Janowska[a]*

[a] Institut de Chimie et Procédés pour l'Énergie, l'Environnement et la Santé (ICPEES), CNRS UMR 7515-University of Strasbourg, 25 rue Becquerel 67087 Strasbourg, France

*Corresponding author: tel: +33 3 68 85 26 33, e-mail: janowskai@unistra.fr

Few layer graphene/carbon (FLG/C) composites are prepared directly via the rapid and simple exfoliation of expanded graphite in the presence of carbon based natural precursors (i.e. protein, polysaccharide) in water, followed by carbonization process. Several parameters such as nature of C-precursor, FLG/C ratio and carbonization conditions (gas, temperature) are modified in order to optimize the morphology, composition and porosity of FLG/C and thereby investigate their impact on gravimetric and volumetric capacitance, their stability and contribution of pseudocapacitance (Ps) vs. double-layer capacitance (DL). Few composites exhibit extremely high capacitance considering their low BET- surface area ranging in 130-260 $m^2/g$. The highest gravimetric and volumetric capacitance of 322 F/g and 467 $F/cm^3$ respectively (0.5 A/g); and energy/power performance is reached for FLG/C:1/2, synthesized from graphite-bovine serum albumin(BSA). Despite relatively high theoretical pseudocapacitance contribution of 69% (1.1V), this sample shows also high capacity retention at high current density and elevated energy -to- power densities. The overall great capacity performance is attributed to the high electrochemical surface area from combined structural features: ultramicroporosity, FLG alignment with high accessibility of FLG edges and elevated packing density. The transport limitation enhances however at higher scan rate (> 100 mV/s).

Keywords: ultramicroporosity, graphene edges, vertically aligned graphene, volumetric capacitance, graphene composites, pseudocapacitance

1. Introduction

The need in global usage of advanced technological devices aiding to human life in various fields has initiated in search for alternative renewable energy production, such as electricity, which in turn has created a surge in research of energy storage systems. The latter ones are traditionally based on batteries or capacitors, while the most recent development concerns the supercapacitors, the systems that hopefully can bridge the gap between relatively expensive and short-life time batteries and capacitors. Supercapacitors are known to work on two-different mechanisms, i) electrical double layer capacitance (EDLC, DL) relating to ion adsorption on high surface area, porous surfaces and, ii) pseudocapacitance (Ps) via slower chemical reversible faradic reactions. The major stepping stones in supercapacitor electrodes developments are carbon based materials including weakly graphitized activated carbon and graphitic carbons (graphene, carbon nanotubes/nanofibers), in which the Ps contribution from redox reactions is usually not negligible especially in activated and/or functionalized carbons and worthy to be considered[1]. Only recently, few works on Ps contribution investigations, in N-doped RGO [2] or graphitized carbons derived from aromatic cross-linked polymers [3], have been reported. The two ways to store energy entail in general trade-off in power and energy performance, i.e. good power and limited energy densities are measured for fast and reversible DL, while higher energy and lower power densities come from slower kinetically redox reactions [4, 5]; and for example the high capacitance and significant energy of highly heteroatom-doped graphene could be obtained (629 F/g at 0.2A/g) but at power density of only 140 W/kg [6].

The initially reported theoretical capacitance of graphene with theoretical surface area of 2675 $m^2/g$ [7], 21 $\mu F/cm^2$ (550 F/g), has been rectified due to the graphene edges contribution. It was found that the theoretical specific capacitance of graphene edges is 4 orders of magnitude higher than the one of graphene planes [8], and for instance, the minimum areal capacitance for edges-oriented graphite reaches 50-70 $\mu F/cm^2$ [9]. The « common » graphene based electrodes including graphene films and reduced graphene oxide (rGO) exhibits usually quite high power density resulted due to relatively high surface area and electronic properties (high conductivity, rapid charge transfer) but they attain moderate capacitance of c.a. 200 F/g [10-12]. Higher values, 349 F/g (2 mV/s), 604F/g (3mA) and 253 F/g (1A/g) were obtained for corrugated/prostruted/perforated graphene/rGO nanorribons/nanomeshs respectively [13-15].

One of the main reason of lowered experimental capacities, far from the theoretical one, is an excessive inter-layer π-π stacking of graphene sheets [16]. To overcome this drawback a formation of graphene aerogels/hydrogels, often by cross-coupling reactions are applied. High surface area, usually between 1000-3000 m$^2$/g and abundant macropores in aerogels result in high specific gravimetric capacitance (222F/g (1A/g) and 325 F/g (1A/g) [17, 18] but in turn entails poor specific volumetric capacitance, $C_{vol}$, due to a low packing density [18, 19]. To increase a density of the active materials some efforts have been undertaken, e.g. converting of graphene hydrogel by capillary pressure and passing from 0.13 to 1.3 g/cm$^3$ yield 255,5 F/cm$^3$ at 0.1 A/g [20], compressing nanoporous N-doped graphene gives 300 F/cm$^3$ (0.6 A/cm$^3$) that is 15-20 times higher than un-compressed sample [21], evaporation-induced drying of graphene hydrogel resulting in 376 F/cm$^3$ [22]. The ultrahigh volumetric capacitance has been reported for densely packed fluorine and nitrogen co-doped carbon microspheres, 521 F/cm$^3$, while for solely N-doped it drops to 262 F/cm$^3$ [23]. Other approach to avoid graphene aggregation deals with a growth and application of high surface area vertically aligned graphene nanosheet arrays benefiting from accessible graphene edges [24-28]. High gravimetric capacitance of such arrays can be reached but still the arrays have large mesopores (intersheets distance) and consequently low volumetric performance. Also sophisticated and relatively heavy techniques are applied to get aligned graphene/FLG, i.e. plasma-enhanced chemical vapor deposition technique PECVD [26], microwave plasma torch (MPT)-PCVD [28], CVD [25], sputtering [29], antifreezing-template-assisted drying [30] or strong magnetic field in resin composites [31]. The milder conditions were just reported for bi-directional freeze casting assisted drying [32]. Only recently, a direct anodic exfoliation of graphite into aligned rGO were reported but volumetric capacitance for graphite-rGO was of only 3.9 F/cm$^3$ in 6 KOH [33]. The alignment of graphene or other 2D materials showed to provide as well other important advantages: the use of high voltage cells in organic liquid electrolytes of the MPT-PCVD grown aligned graphene [28] and of the mesoporous carbon with edge-free graphene walls [34]; the benefit from the directional transport on the capacity efficiency (thickness independent) was also demonstrated in aligned MXenes [35].

The activated porous carbons are another large family of commonly used capacitive electrodes. They are in general prepared from the synthetic or natural precursors such as polymers [36], synthetic organic molecules [2], sugars or directly from bio-mass [37](fruits, fruits shells, etc). In term of specific capacitance, they can reach at low current densities the values such as 374 F/g (0.1 A/g) [2]. The similar capacity performances have been observed in several graphene-carbon composites, i.e. the graphene-carbon composites reached the capacitance of 324 F/g at 0.3 A/g in 1M $H_2SO_4$ [38]. The composite of polypyrrole and graphene has shown capacitance of 326 F/g at 0.5 A/g (250 F/g at 2A/g) (obtained by electropolymerization) [39]. The composite, sugar-derived carbon/graphene has shown 273 F/g at 0.5 A/g in 6M KOH [40]. Still other examples deal with the use of carbon nanotubes (CNTs) as graphene spacers: 211 F/g at 0.5 A/g [41], or double spacers CNTs and NPs reaching in rGO/CNTs/NiO composite the capacitance of 1180 F/g at 1A/g [42]. High gravimetric and volumetric capacitance were obtained for hybrid film of GO/NaCl/ urea at 1A/g in 1M $H_2SO_4$ in symmetrical cell (425 F/g and 693 $F/cm^3$, respectively) [43]. High volumetric capacitance of 364 F/g was obtained also for multilayered rGO/polymer architecture [44].

Apart from some issues described above there are other ones linked to the preparation of electrode materials for high scale applications. Most forms of graphene used is in fact the reduced graphene oxide, the synthesis of which, including reduction, is relatively long and harsh process [45], while the graphene monolayer doesn't solve the issue of expensive scaling and difficulty to be obtained at large scale by exfoliation methods. The activated carbon electrodes originating from the natural sources (fruits) suffer in turn from a weak control of their final structures.

Herein we present the electrodes that are composites of few layer graphene (FLG) and carbon (FLG/C), obtained by *in*-situ exfoliation of graphite in the presence of natural bio-compatible C precursor, followed by carbonization process. The preparation method makes no room for waste in the entire process, while water is used as the exfoliation medium, making the approach attractive from the economic and environmental point of view.

The C from the natural precursors (albumin/maltodextrin), as described below plays the role of spacer for FLG, inducing interesting FLG-C compact arrangement, also facilitating vertical alignment, and nanoporosity. The electrochemical measurements are performed in three-electrode system in 0.5M $H_2SO_4$ electrolyte. The DL and Ps capacities contributions as well as electrochemical performance – structure relation are investigated.

2. Experimental

2.1. Synthesis of FLG/C composites

Expanded Graphite (EG) was purchased from Carbon Lorraine company, Bovine Serum Albumin (A) and Maltodextrin(M) were purchased from Sigma-Aldrich and MyProtein company respectively.

General approach: x mg of bovine serum albumin (and y mg of maltodextrin) was first solubilized in 200 ml of deionized water by stirring for about 15 min. Then z mg of EG was added to the solution under constant stirring at 700 rpm for about 15 min, such that all of added EG is visibly wet and mixed with the solution. Then the mixture is subjected to simultaneous stirring and sonication (top ultrasonicator) in a cooling water bath for 2 hours. After the completion of the reaction, the obtained suspension is dried using freeze-drying method for 24 hours at -80°C. The sample is then carbonized in a quartz reactor at 600°C for 2 h under $NH_3$ atmosphere. The heating and cooling of the reactor is done under Ar atmosphere. The sample thus obtained is used to fabricate electrodes for electrochemical tests.

Table below includes the compositions of the composites before (FLG/A/M) and after (FLG/C) carbonization.

| FLG/C (final composition/after carbonization) | FLG/hHLB (FLG/A/M, initial composition/before carbonization) |
|---|---|
| 1/2 | 0.5/4/0 |
| 1.4/1 | 1/3/1 |
| 1/1.2 | 1/1/3 |
| 1/1.6 | 0.5/3/1 |
| 1/2.3 | 0.5/1/3 |
| 1/2.5 | 0.25/3/1 |

Table. 1 Final (after carbonization) and initial (before carbonization) compositions of FLG-C.

2.2. Preparation of electrodes

For fabrication of electrodes, circular glassy carbon (GC) of 5mm diameter obtained from HTW Hochtemperatur-Werkstoffe GmbH, Germany was used. Each of the GC was initially polished with sand paper on one side to have a mirror-like reflection surface. Further, it was sonicated in ethanol for five minutes and washed three times with ethanol to make sure it is free from surface contaminants. The ink for the GC was made with a suspension of FLG/C samples and ethanol with a conc. of 1mg/ml. To this, 1% by vol. of Nafion (5% by vol. in ethanol) was added. It was sonicated for 20-30 min. to obtain a homogenous ink. The ink was then drop-casted on the GC using a pipette of 10-100 µl precision. Each cast was generally of 10-20 µl. Considering the visual surface coverage of the GC, each electrode was made of 100-140 µg of the composite. After drop-casting, the electrode was air-dried in oven at 60°C for 15 min. to make sure there was no ethanol left. Then, it was used for electrochemical test.

2.3. Characterization tools

XPS analyses were performed with a MULTILAB 2000 (THERMO) spectrometer equipped with Al K anode (hv= 1486.6 eV). For deconvolution of spectra, the CASA XPS program with a Gaussian−Lorentzian mix function and Shirley background subtraction was employed.

Transmission electron microscopy (TEM) was carried out on a Topcon 002B - UHR microscope working with an accelerated voltage of 200 kV and a point-to-point resolution of 0.17 nm. The sample was dispersed in ethanol and a drop of the suspension was deposited onto a holey carbon coated copper for analysis.

Scanning electron microscopy (SEM) analysis was carried out on a Jeol JSM-6700F working at 3 kV accelerated voltage, equipped with a CCD camera.

TGA analyses were carried out on TA instrument SDT Q600, the rate of the heating was fixed at 10º per min under air flow.

BET surface area and porosity analysis were carried out on Micrometrics ASAP 2420 instrument, by the adsorption of $N_2$ at 77K. Each sample were measured few times at different conditions of time and temperature degassing as well as different interval equilibration time. The microporous surface area was obtained by Dubinin-Radushkevich method.
For the analysis of microporosity between 0.5 and 0.9 nm a dosage amount of $N_2$ was decreased to 2-6 $cm^3$/g STP with an max. equilibration time of 2-3h, and the tubes were calibrated prior to the measurements.
The pores size distribution was obtained by BJH method (comparing the desorption and adsorption branches and excluding eventual artefacts [46].
The microporosity size distribution were estimated in accordance to Horvath-Kawazoe method.

BET surface area and microporous surface area analysis for sample FLG/A: 1/2 was also carried out on Micrometrics ASAP 2420 instrument by adsorption of $CO_2$ at 273 K. The pore size distribution was obtained for this sample by DFT method.

The densities of the prepared FLG/A/M composites were measured using a liquid pycnometer. For each sample, the average of 3 measurements were used.

2.4. Electrochemical measurements

The electrochemical measurements were carried out in a three-electrode cell setup using a Biologic SP-300 system. The three-electrode setup had GC with sample material as working electrode, platinum wire as counter electrode and Mercury Standard Electrode (MSE) as reference electrode. All the measurements were performed in 0.5M $H_2SO_4$ electrolyte at potential window of 1.1 V. The electrolyte was first degassed with $N_2$ for at least 20 min before starting any experiments.

The specific capacitance was obtained from galvanostatic charge/discharge measurements according to the equation:

$$C_{GCD} = \frac{It}{m.V} \ (Fg^{-1})$$

where, $I(A)$, $t(s)$, $m(g)$ and $V$ represent the discharge current, discharge time, mass of active material and discharge voltage (The slopes from the linear fit of linear regions from the discharge curves were used).

The voltametric capacitance, $C_{CV}$ was calculated from the CV curve according to the formula:

$$C_{CV} = \frac{CV \ Area}{2 \ m \ s \ \Delta V} \ (Fg^{-1})$$

where, $s$ represents potential scan rate at 5mV/s and $\Delta V$ the potential window of 1.1 V.

The specific energy density (SE) and specific power density (SP) was calculated according to the equations:

$$SE = \frac{1}{2}C_{GCD}\Delta V^2 \ x \ \frac{1000}{3600} \ (Wh \ kg^{-1})$$

$$SP = \frac{SE}{t} \ x \ 3600 \ (W \ kg^{-1})$$

The volumetric capacitance was obtained according to the equation:

$$C_{vol} = C_{GCD} \times d \ (F\ cm^{-3})$$

where *d* represents the density of active material.

3. Results and discussion

3.1 Morphology/composition of FLG/C

The first step of FLG/C composites' preparation in water relates to the *in*-situ occurring exfoliation of EG that is possible due to the presence of large natural bio-surfactants with high hydrophilic-lipophilic balance (hHLB) and ultrasonication/mixing action[47, 48], FLG/hHLB. In the present study two hHLB systems, bovine serum albumin (A) and maltodextrin (M), are used for exfoliation process and as precursors of the carbon that is next formed via carbonization treatment yielding the composites (FLG/C). The composites are denoted as FLG/A and FLG/A/M, since all composites were formed from the mixture of FLG and A or FLG, A and M, with different ratio (Table 1). On the contrary to a simple exfoliation, the amount of used hHLB vs. FLG (also mixed A/M) is much higher in order to get sufficient final C content (300-1600% compared to 10% in a simple exfoliation). This amount decreases of course in significant way after carbonization while the final ratio of FLG/C is found by TGA analysis and e.g. resulting from FLG/A/M of initial ratio of 1/3/1 the final composite is defined as FLG/A/M:1.4/1, other one, the sample FLG/A/M:1/1.2 results from FLG/A/M of initial ratio of 1/1/3 (Fig. 1a). According to the TGA analysis higher amount of C is found after carbonization in the composites containing an excess of M vs. A. Most of the thermal decomposition of hHLB into C in the composites are performed under $NH_3$ at 600°C and are called simply carbonization, however it entails N-doping as well and the formation of new N-type groups can be observed by XPS spectroscopy. For the reason of comparison few selected composites are subjected to carbonization under Ar or temperatures lower than 600°C.

The exemplary C1s and N1s XPS spectra of FLG/A/M: 1/3/1 before and after carbonization in Ar and $NH_3$ are presented in fig. 1 b and c respectively. Table 2 illustrates the relative % of C, N and O obtained from XPS and elemental analysis for selected samples. As one can see the C1s spectra (Fig. 1b) change totally after carbonization from the $sp^3C$ and $sp^3C$-heteroatoms rich-signature at higher binding energy (BE) values related to the presence of hHLB(A/M) into $sp^2C$ rich-signature at ~ 284.5 eV from FLG. The significant modification of N1s spectra after carbonization (Fig. 1c) includes a disappearance of the N-peaks from albumin (essentially amide groups) at higher BE and appearance of new peaks at lower BE from newly formed groups. A deconvolution of N1s spectra, exemplary from FLG/A: 1/2, Fig. 1e, indicates the presence of four type of peaks attributed to the pyridinic-N at 398.4 eV, pyrrolic/pyridone-N at 400.3 eV, quaternary- N at 401.5 eV and N-oxide (N-X) at 403.3 eV [49]. The fig. 1f illustrates the % contribution of the N-containing groups in the most important samples and the one carbonized under Ar. In all composites the pyridinic and pyrrolic/pyridone N groups are in high excess compared to the rest and according to the literature these are the groups providing a pseudocapacitance from proton exchange [2]. In the composite subjected to $NH_3$ assisted doping a total % of N reaches c.a. 7-8%. The much lower % of N observed for Ar-treated sample is because of a doping effect from the decomposition of albumin. Consequently, the amount of N is poorer in the sample with initially high M content (FLG/C:1/1.2). We do not investigate in details the oxygen effect since its % is relatively low compared to N, while commonly only quinone-type groups can undergo redox reaction efficiently at low pH [2]. Likewise, the amount of sulfur that could remain in the composites from the albumin (e.g. cysteine) is hardly detected and consequently not considered for the investigations.

The morphological investigation by TEM micrography of carbonized FLG/C composites reveals that obtained FLG flakes have few microns' lateral size and 5-10 number of sheets, in average (Fig. 2 a, b). In the carbon layer adsorbed over FLG surface the mesopores of c.a. 10-15 nm can be locally observed (Fig. 2c). This carbon derived from hHLB covers the surface of FLG flakes and at some extend links FLG flakes (Fig. 2d). Some hHLB spheres are observed before the carbonization, that constitute an excess and is not adsorbed over FLG surface (Fig. 2e).

An interesting difference was observed within the structures of the composites with variable A/M content at micro/macro-scale. The FLG/hHBL containing solely an albumin or its excess (vs. M) exhibits a strong tendency to form vertically aligned patterns. Fig. 3 (a,b) are SEM micrographs of FLG/A: 0.5/4 after freeze-drying. The degree of alignment increases with A content and is almost absent in FLG/A/M: 1/1.2 (in. 1/1/3). On the contrary, the samples containing higher excess of maltodextrin demonstrates optically more typical porous, foam like structure. (This difference can already be observed in separately carbonized A and M alone, optical photos, Fig. 1a-c SI). The vertically aligned FLG flakes in FLG/A (0.5/4) (fig. 3

a,b) form locally the hexagonal network in the horizontal plane (Fig. 3b). We assume that the observed tendency of alignment in A-rich FLG/hHLB composites are caused first of all by hydrogen bonding enhanced at some points by van der Waals interactions. The part of albumin adsorbed on FLG surface is denaturized and exist in secondary linear structure. The hydrogen bonding responsible for natural tridimensional conformation of albumin is destroyed during the ultrasonication, while new hydrogen bonds are formed between adsorbed linear chains of A on FLG and water, that next freezes and sublimes during the freeze-drying process. The linear A chains are uniformly adsorbed on FLG surface giving an access to homogenous and densely localized hydrogen bonding groups. One can find in the literature only recently reported freeze dried GO that alignment symmetry could be modified by addition of different additives such as ethanol, chitosan or ethanol [32]. In spite of the fact that the structures are somehow affected by carbonization process, less dense but still aligned patterns are observed (Fig. 3c) (It is worthy to note that it was possible to handle and manipulate the carbonized A-rich composites with some limitation but no difference was observed in specific gravimetric capacitance if compared to dispersed and dried composites' electrodes). Even though the composites are dispersed for the preparation of electrodes, a dense compact arrangement is again observed in dried electrode material Fig. 3d, e: side view and Fig. 3f: top view (locally, small extra part of less dense FLG/C splits out of the aligned patterns of Fig. 1 d SI). In the in-set of Fig. 3f in top-view, we can see some edges of the protruding FLG flakes. Such arrangement can be substantiated by a diffusion limited aggregation (DLA), the process occurring during the solvent evaporation. Such evaporation-induced aggregation are either unwanted in colloidal science or it brings interesting structures and e.g. in the case of FLG, the flat conductive fractal-like networks assemblies were obtained through DLA [50]. Several factors play a role in how DLA process runs and what kind of arrangement comes out and deeper studies are necessary to describe with certitude herein assembling process (are undergoing).

Table 3 in Fig. 4a groups the values of specific surface area $S_{BET}$, and $S_{micr.}$ contribution from micropores obtained by Dubinin-Radushkevich method, for the most important selected composites. All composites show low $S_{BET}$, first, due to the relatively important thickness of FLG, and second and more significantly, due to the probable partial π-π interlayer re-stacking of the FLG during degassing process prior to the BET measurements. Relatively low $S_{BET}$ of the composites is not surprising since we expect that the produced FLG flakes include minimum 5,6 graphene sheets in average, according to previously reported method for the exfoliation of graphite (EG) [47]. The exfoliation efficiency depends indeed on the nature of hHLB and hHLB: EG ratio that vary strongly within the present composites family. In addition, no separation step that can remove the thicker FLG flakes (sedimentation) after the exfoliation is applied here. Consequently, considering a theoretical value of surface area of graphene of 2675 m$^2$/g, the obtained here surface values indicate the presence of 6 to 20 sheets in average within the FLG flakes depending on the composite. All composites show Type I isotherms and are predominantly microporous Fig. 4b. Some mesopores are detected through the pore size analysis that could correspond to the pores ~ 10 nm observed by TEM. The highly microporous (ultramicoporous) character can be observed also by the behavior of the N$_2$ adsorption and desorption isotherms, where necessary efforts need to be undertaken to choose the appropriate conditions for N$_2$ adsorption (equilibration time, degassing time and temperatures) [46, 51, 52]. The desorption isotherms do not completely overlap at lower pressure range. This suggests the presence of slit type pores, in which the lateral interaction between the adsorbed molecules is larger than the interactions between the C-covered FLG flakes during the adsorption and the width of slit pores (space between the flakes) increase. The materials are not rigid, what additionally can impact sorption process. The second adsorption analysis at lower N$_2$ speed supply, Fig. 4c, allowed to detect the very small micropores between 0.5 and 0.9 nm, ultramicropores, with relatively large distribution but for some samples a significant excess of pores with diameter around 0.5 nm (and/or below) were detected. These ultramicropores can be attributed to the slit pores created between aligned C-covered FLG flakes. For the reason of comparison, the adsorption was also performed by CO$_2$ uptake for one sample, FLG/A:1/2 (the method being more appropriate for the micropores < 1nm). The analysis resulted in relatively comparable S$_{BET}$ and S$_{micro}$ with N$_2$ adsorption values, while distribution of micropores size via DFT method confirmed the excess of pores of around 0.47 nm, fig. 4d.

3.2 Supercapacitance of FLG/C composites

The electrochemical performance of FLG/C composites were measured in 0.5M H$_2$SO$_4$ aqueous electrolyte in three electrode set-up and at a potential window of 1.1 V. Fig. 5 illustrates the cyclic voltammograms at 20 mV/s (a), galvanostatic charge-discharge curves (GCD) at current density of 1A/g (b), a graph with specific gravimetric capacitance values and their retention at higher current densities (0.5 A/g to 20A/g) (c) and cyclic stability test of 6000 cycles (capacitance retention) at 20 A/g for the best sample (d).

Five composites were investigated (the composites treated under Ar and/or under lower temperature (400°C) exhibited lower electrochemical performance compared to the ones carbonized/N-doped under $NH_3$ at 600°C and consequently they will be not discussed here). In all composites the symmetrical CV curves and triangular shapes of the GCD for all composites reveal good reversibility behavior, also, the CV curves of few samples show quasi-rectangular shape.

The CV current density first increases, next stabilized and finally decreases systematically along the potential shift towards positive values. The potential window of the stable current density corresponding to double layer capacity contribution varies however with the samples being low in sample poor in FLG e.g. FLG/A/M:1/1.6 and relatively large in FLG/A/M/:1.4/1. The highest capacity, 322 F/g at 0.5 A/g, is measured for FLG/A:1/2 (in. 0.5/4), the composite initially consisting of albumin hHLB solely. Two other best capacitances at 0.5 A/g belong to the composites with similar initial ratio to the best sample of FLG/C, i.e. 208 F/g for FLG/A/M: 1/1.6 (in. 0.5/3/1) and 187 F/g for FLG/A/M:1/2. (in. 0.5/1/3). Likewise, the FLG/A/M:1/1.14 originating from FLG/A/M:1/3/1 demonstrates higher capacity than FLG/A/M:1/1.2 originating from FLG/A/M:1/1/3. The composites with high amount of M show lower capacities compared to their A-rich counterparts. There is no correlation between capacitance of the composites and $S_{BET}$. The three composites exhibiting the highest but different capacitances at 0.5 A/g have similar $S_{BET}$. The composite with highest $S_{BET}$ ( FLG/A/M:1/2.5) show very low capacitance (F ≈ 22 F/g, 0.5 A/g). Simply, also all the $S_{BET}$ values obtained for FLG/C composites are extremely low considering very good capacities demonstrated by the composites. This inconsistency comes from the fact that the $S_{BET}$ does not necessarily reflect the effective surface area accessible by electrolyte ions (E-SSA) [53], especially if ultramicropores are present in excess. The "traditional" and "basic" figures of merit for design of high capacitance, i.e. high specific surface area and sufficiently large pores diameter, were shown to collapse in the case of "nanoporous" materials. In carbide derived carbon, the capacitance decreased with lowering the pore size to the size lower than twice the solvated electrolyte ions, then it increases again for the pores below the size of solvated ions, below 1 nm [54]. It seems that the optimal initial ratio of hHBL vs. FLG is like 8 to 1 (FLG/A:0.5/4, FLG/A/M: 0.5/3/1). The increase or decrease of hHBL content results in lower capacity values. The capacitance values obtained from A and M alone prepared at the same conditions as the FLG/C shows very low capacitance of c.a. 7 F/g. Such small value suggests that most of the measured capacitance in the composites comes from FLG, maintaining in mind, that the presence of minimum amount of C that is necessary to make the spacer for FLG flakes. From the other hand the FLG flakes permit first for A/M adsorption and consequently better dispersion of C. The additional capacitance investigations were performed for FLG/A/M: 1.4/1 (in. 1/3/1 carbonized under Ar or under $NH_3$ but at lower or higher temperature (200, 400 and 800°C). All these samples showed much lower electrochemical performance compared to the ones carbonized under $NH_3$ at 600°C (SI).

The highest capacitance retention at higher current density (up to 20 A/g) is observed for the composites rich in FLG, fig. 5c, confirming a great ability of graphene/FLG to sustain high current density. (The additional measurements of stability was performed for doubled mass of FLG/A samples, i.e. 1 mg/cm$^2$, SI). Still, the stability is very good, while the capacitance is more than twice lower probably due to the increase of the samples thickness with related slowed transport through the sample film and /or modification of the FLG alignment/density). The excellent retention is measured for the FLG/A:1/2, its capacitance drops from 322 F/g at 0.5 A/g to 222 F/g at 20 A/g, i.e. ~ 30% of drop. Slightly higher drop of capacitance of 37% is measured for FLG/A/M: 1.4/1 (in. 1/3/1), from 121 F/g to 76 F/g, respectively, while a significant capacitance reduction is successively observed for the samples containing higher amount of C, i.e. ~ 88% for FLG/A/M : 1/1.6 (206 F/g at 0.5A/g to 25 F/g at 20 A/g) and ~ 61 % for FLG/A/M : 1/2.3. In the initially M-rich sample (in. FLG/A/M : 1/1/3), despite relatively high amount of FLG, a drop of capacitance is relatively high (~ 65%), and the capacitance is very low, 54 F/g at 0.5A/g. Fig. 6 illustrates also CV curves at different potential scan rate up to 500 mV/s (a-c) and GCD curves for current densities between 0.5 to 20 A/g for three composites (d-i) , FLG/A: 1/2, FLG/A/M:1/1.6 (excess of A) and for the one containing initially an excess of M, FLG/A/M:1/2.3 (excess of M). As one can see with the increase of potential sweep, the CV curves exhibit increasing features of pseudocapacitance in FLG/A/M:1/2 and especially in FLG/A/M:1/2.3. It has to be underline that the composites systems above 100mV/s start to adapt a leaf-like shape with the fastest distortion from square-like in FLG/A/M: 1/2.3. This behavior at high scan rate, far from "ideal" supercapacitor, can be probably associated with electrical transport limitation as well as the electrolyte transport limitation, i.e. resistance in the ultramicroporous that limit effective, accessible surface area [55, 56].The impact of ultramicropores structures seems to be significant, the highest $S_{micro}$ was measured in FLG/G/A:1/2.3; consequently the interfacial DL capacitance does not have enough time to form at high fast scan rate.

In order to investigate the contribution of $C_{Ps}$ originating from redox reactions vs. $C_{DL}$ we applied the method recently adapted for carbon based materials [2, 3] from metal oxides [57]. Accordingly, the total voltammetric charge $Q_T$ is obtained by plotting $1/q$ vs. $v^{1/2}$ and extrapolation of the tendency curve to $v = 0$, while $q_{DL}$ by plotting $q$ vs. $v^{-1/2}$ and extrapolation of charge (tendency curve) to $\infty$, figure 7 a, b. The corresponding $C_T$ and $C_{Ps}$ values are next obtained by dividing $q_T$ and $q_{DL}$ by the potential window (1.1V), while $q_{Ps}$ and/or $C_{Ps}$ can be consequently achieved by differencing corresponding T and DL values.

The obtained voltammetric capacitance values: maximum total $C_T$, $C_{Ps}$, $C_{DL}$ and $C_{Ps}$ % are presented in table 4, fig. 7 c, as well as calculated from the CV curves the CV capacitance ($C_{CV}$). The increasing tendency of calculated $C_T$ is consistent with the obtained $C_{GCD,}$ except one sample which further shows very high pseudocapacitive contribution. This latter, pseudocapacitive contribution, is probably a main reason of the observed discrepancies between $C_T$ and $C_{GCD}$ in all composites. The discrepancies increase with increase of $C_{Ps}$%. Due to the relatively slow sweep in CV, the $C_{CV}$ were calculated at 5 mV/s. Likewise, the omitted initial part of discharge curves (GCD) related to IR drop form the internal resistance could be also the reason for the difference between obtained $C_{GCD}$ and $C_{CV}$ for few samples.

The estimated $C_{Ps}$ varies from 68 to 87%. They are quite high even if taking into account the fact that most of the composites has significant amount of C. As mentioned above, if we look however the CV curves as a function of potential sweep, fig. 6 a-c, we clearly find the evolution of $C_{Ps}$ contribution with increasing of V sweep, especially for FLG/A/M:1/2.3, the sample with the initial excess of M. Considering a $C_{DL}$ contribution, the estimated values do not follow obtained $S_{BET}$ or $S_{micro}$ area. We assign this behavior to different edges exposure of FLG sheets, which differs more due to the composites' architecture and not to the edges content. Consequently, despite relatively low $S_{BET}$, the best samples show high $C_{DL}$ contribution, since it origins from the easily accessible aligned flakes with exposed edges but also inter-flakes slit ultramicropores. Let's recall that FLG/A (FLG:1/2) is the one where the alignment of the FLG and consequently the exposition of the edges is maximal. Good accessibility of the FLGs' edges in the composites, exhibiting FLG's alignment tendency (initially A-rich), induce higher electrochemically active surface area. As mentioned in the introduction, the input of the graphene edges into total capacitance was confirmed recently several times [8, 9, 13-15, 58-60]. In the theoretical studies using edges-passivated graphene nanoribbons the effect of edges on quantum and DL capacitances were clearly demonstrated [60]. The graphite containing the KOH-etched pits with very small micropores and exposed edges showed optimal capacitance [61]. It was found that the capacitance decreased with increase of SSA. The exposure of the edges and significant ultramicroporous feature entails also high volumetric capacitance [54, 61], the microporosity maximize a density of ion storage and the exposed edges with superior ability for charge accumulation compared to the basal plane. This is also true for the present FLG/C composites. The observed nanoporosity comes indeed from the space between vertically aligned FLG-C flakes of those the edges which are "necked" towards the electrolyte ions.

Due to the alignment and quite compact structure, most of the composites exhibit very high densities, between 0.44 and 1.65 g/cm$^3$, and consequently high volumetric capacities (Table 3, Fig. 7c), with the best sample showing the volumetric capacity as high as 467 F/cm$^3$. Such high densities for densely aligned FLG-C flakes are expected; they are placed between the low density of graphene (in 3D foam structures) and that of graphite (~ 2.2 g/cm$^3$). Relatively low densities were measured for two composites with higher FLG content (0.88 and 0.44 g/cm$^3$) which can be due to the insufficient C-to-FLG ratio and more hydrophobic character from FLG counterpart; these two composites are the ones with lower M content.

The Ragone plots, based on mass and volume normalized capacitance values show similar tendency within the composites except FLG/A/M: 1.4/1, of which C$_{vol}$ significantly decreases due to the relatively low density, Fig. 7d, e. The three composites with optimal initial ratio of FLG/A/M: 0.5/4, i.e. FLG/A/M: 0.5/1/3, FLG/A/M: 0.5/3/1 and FLG/A: 0.5/4 revealed higher volumetric energy and power densities compare to the gravimetric merits. The most significant change is established for FLG/A/M: 0.5/1/3, for which additionally relatively fast discharging results in very high power density of 0.437 W/cm$^3$ and energy of 0.052 Wh/cm$^3$. The best sample, i.e FLG/A shows 54 Wh/kg at 269 W/kg of gravimetric and 0.078 Wh/cm$^3$ at 0.39 W/cm$^3$ of volumetric energy and power densities respectively. The obtained performance is quite high compared for instance with the results obtained also for one electrode in N-graphene [6].

4. Conclusion

FLG/C composites with different FLG/C ratio were prepared by direct exfoliation of graphite in water, followed by carbonization-N-doping treatment, where C were initially albumin or albumin/maltodextrin. The DL and Ps capacitance contribution are estimated for the composites. Despite very low specific surface area $S_{BET}$, high capacities are obtained for FLG/A/(M) with general ratio of 0.5/4, i.e. FLG/A/M: 0.5/3/1, FLG/A/M: 0.5/1/3 and the best one for FLG/A: 0.5/4. The later one exhibits the gravimetric and volumetric capacity of 322 F/g and 467 F/cm$^3$ at 0.5 A/g respectively and shows superior stability at high current density. The great performances obtained for few composites are the result of their unique architecture; first, a presence of ultramicropores; second, a vertical alignment of FLG-C flakes was obtained in the composites with albumin or albumin excess composites. The vertically aligned FLG/C brings great edges exposure increasing efficient electrochemical surface. Most of the composites have also high to very high packing density, what entails non-negligible volumetric capacitances. The charge and ionic transport limitations are present at scan rates above 100 mV/s especially for the samples with high C content and $S_{micro}$ area.

Author contributions. A. Mohanty carried out most of the experimental part (synthesis, electrochemical measurements, some characterizations), I. Janowska was PI/coordinator of the work and wrote the manuscript.

Acknowledgements

The authors want to acknowledge FRC foundation for FRC- (Solvay) founding. Dr. Sergey Pronkin (ICPEES) is acknowledged for his help with electrochemical set-up and related discussions, Dr. Vasiliki Papaefthimiou for performing XPS-measurements and Dr. Thierry Dintzer (ICPEES) for carrying out SEM microscopy. Dris Ihiawakrim from IPCMS is acknowledged for performing TEM microscopy. Ksenia Parkhomenko (ICPEES) and Camelia Matei Ghimbeu (IS2M) for help with N$_2$ and CO$_2$ gas adsorption measurements, respectively.

Figure1

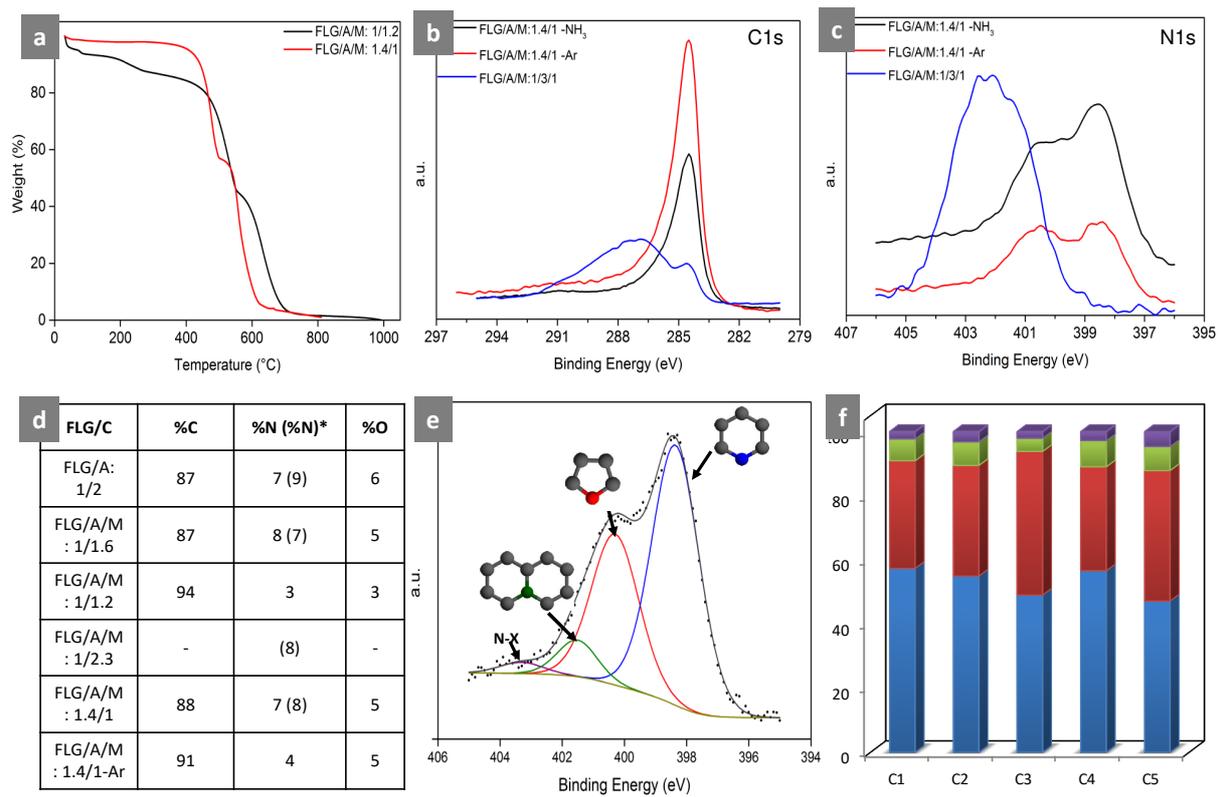

Figure 2

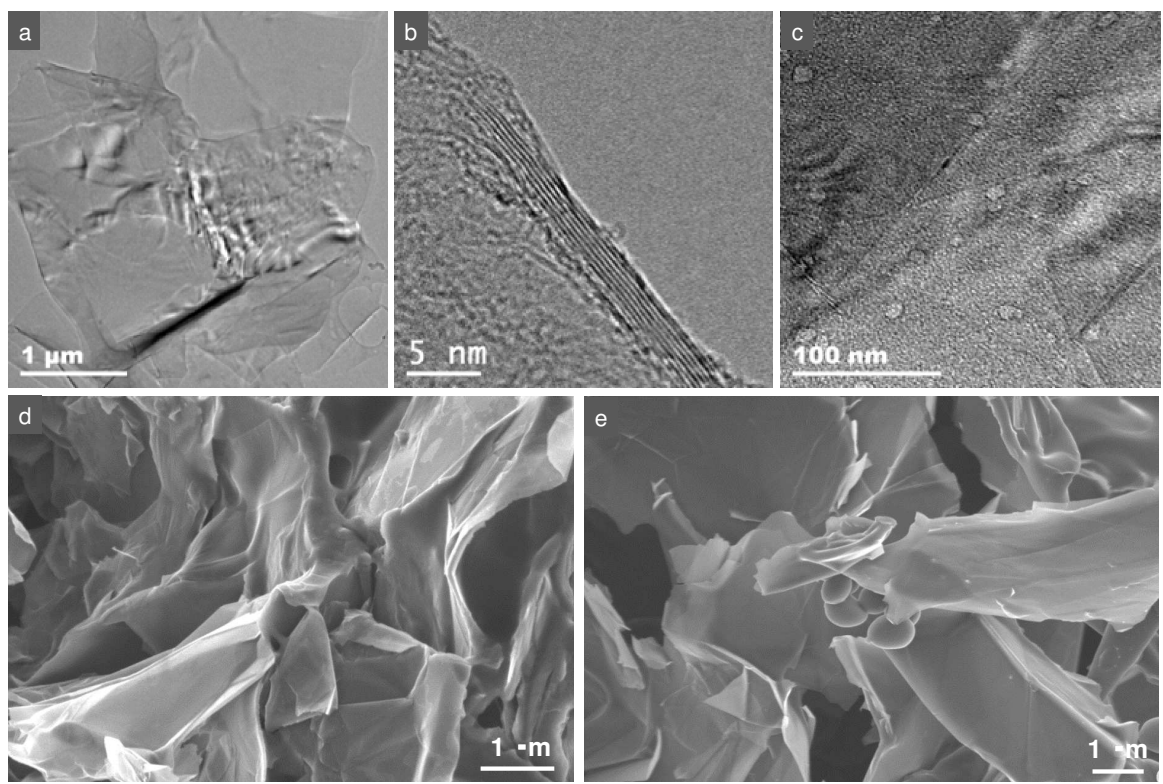

Figure 3

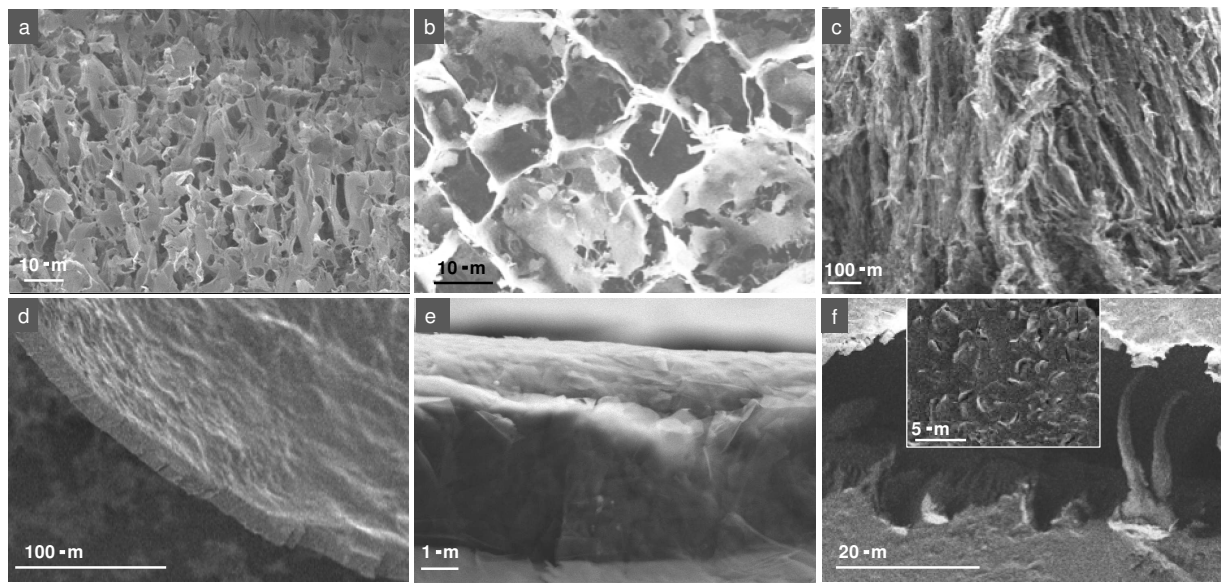

Figure 4

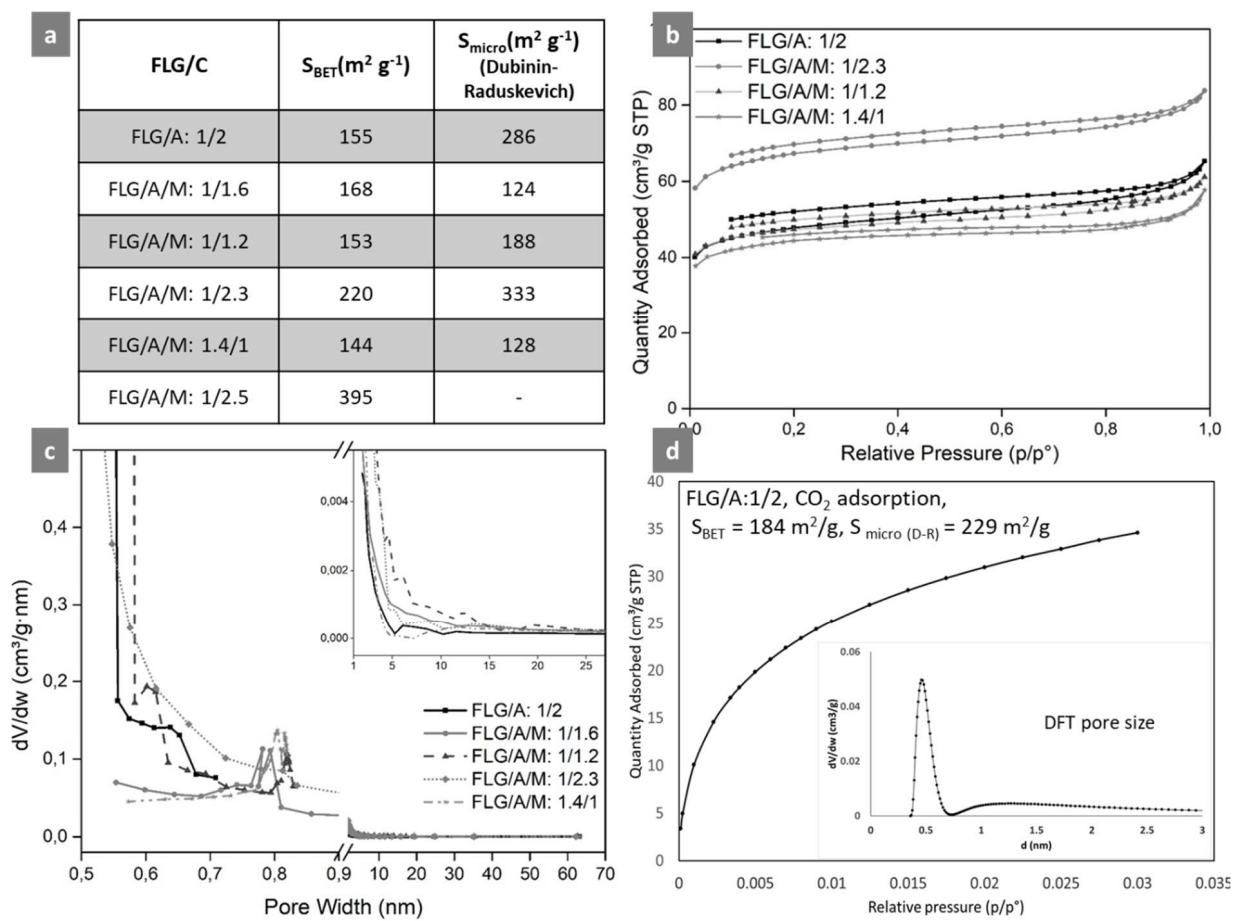

Figure 5

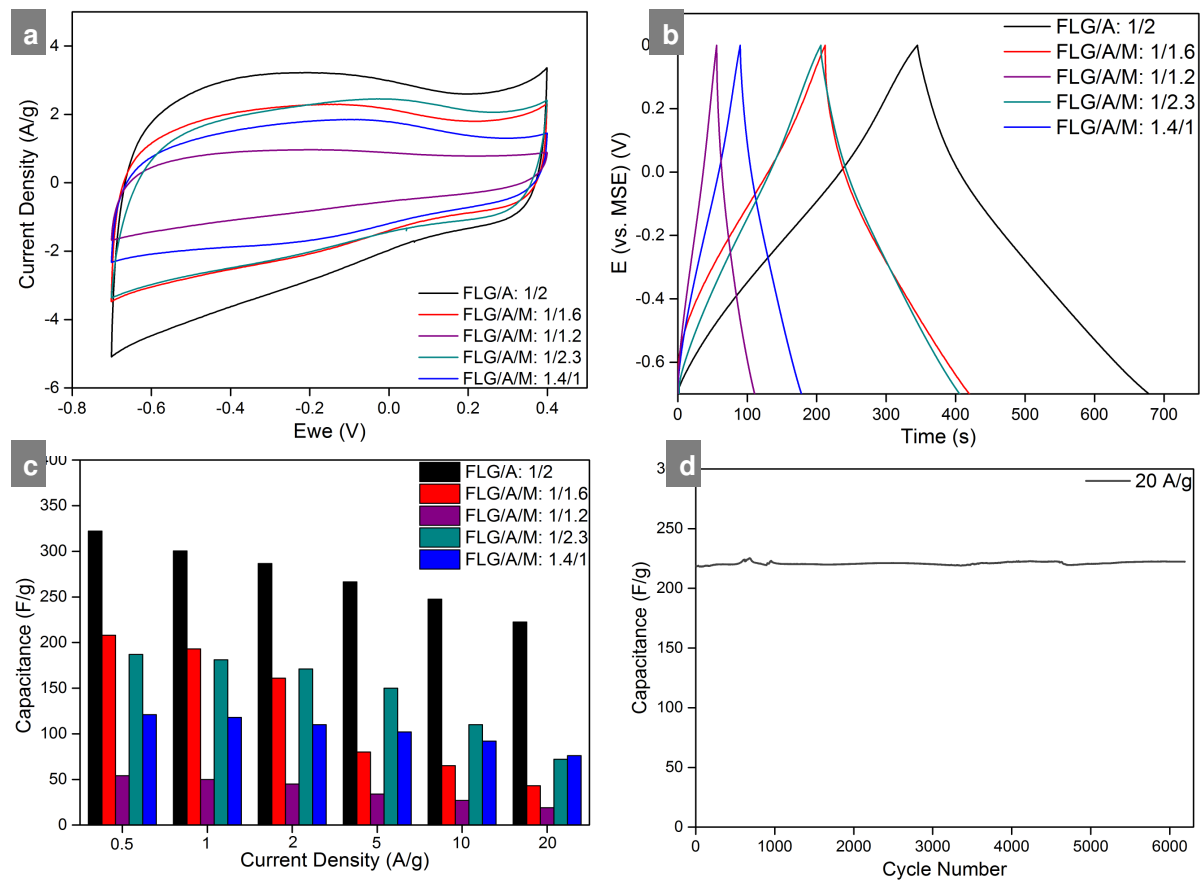

Figure 6

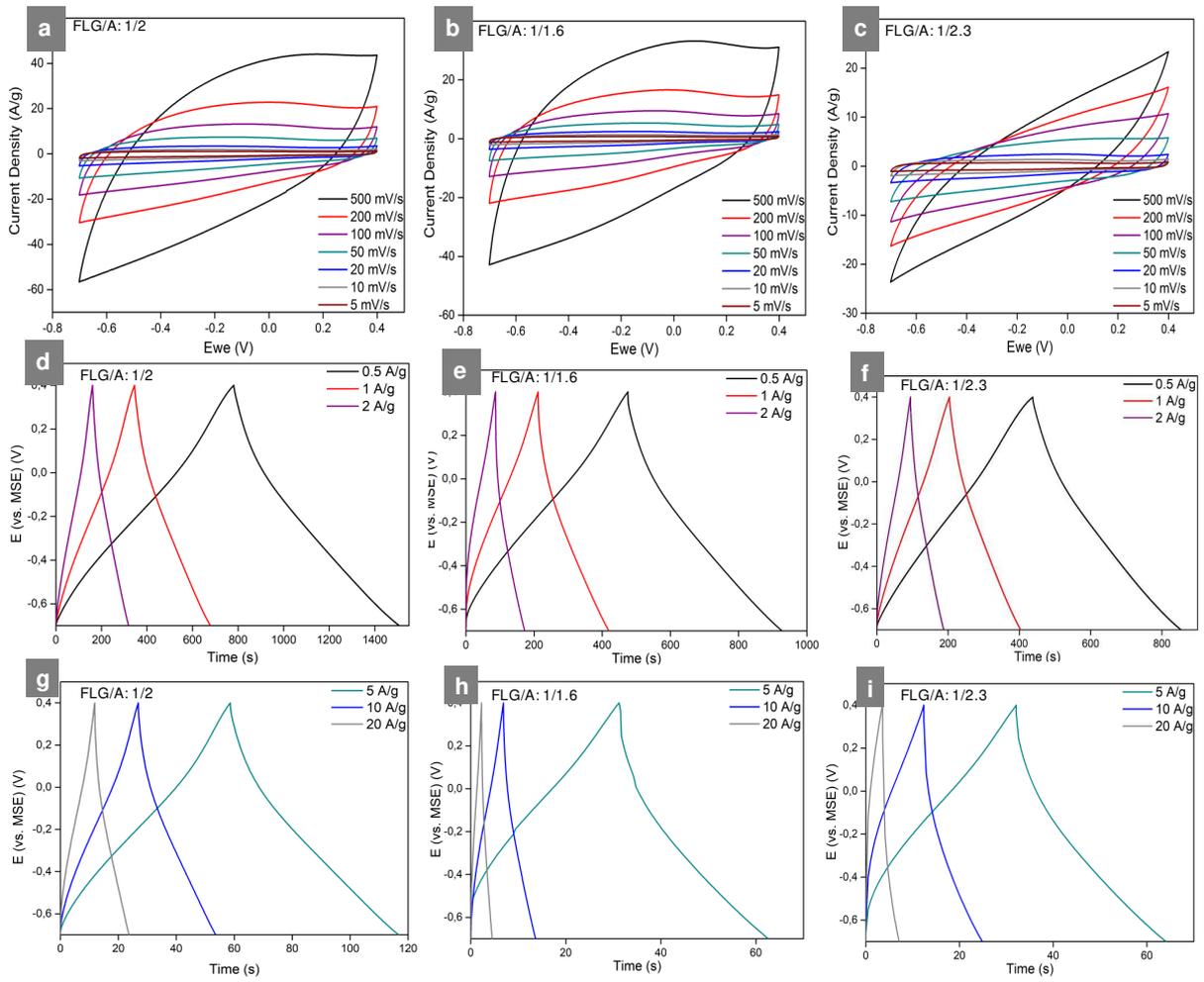

Figure 7

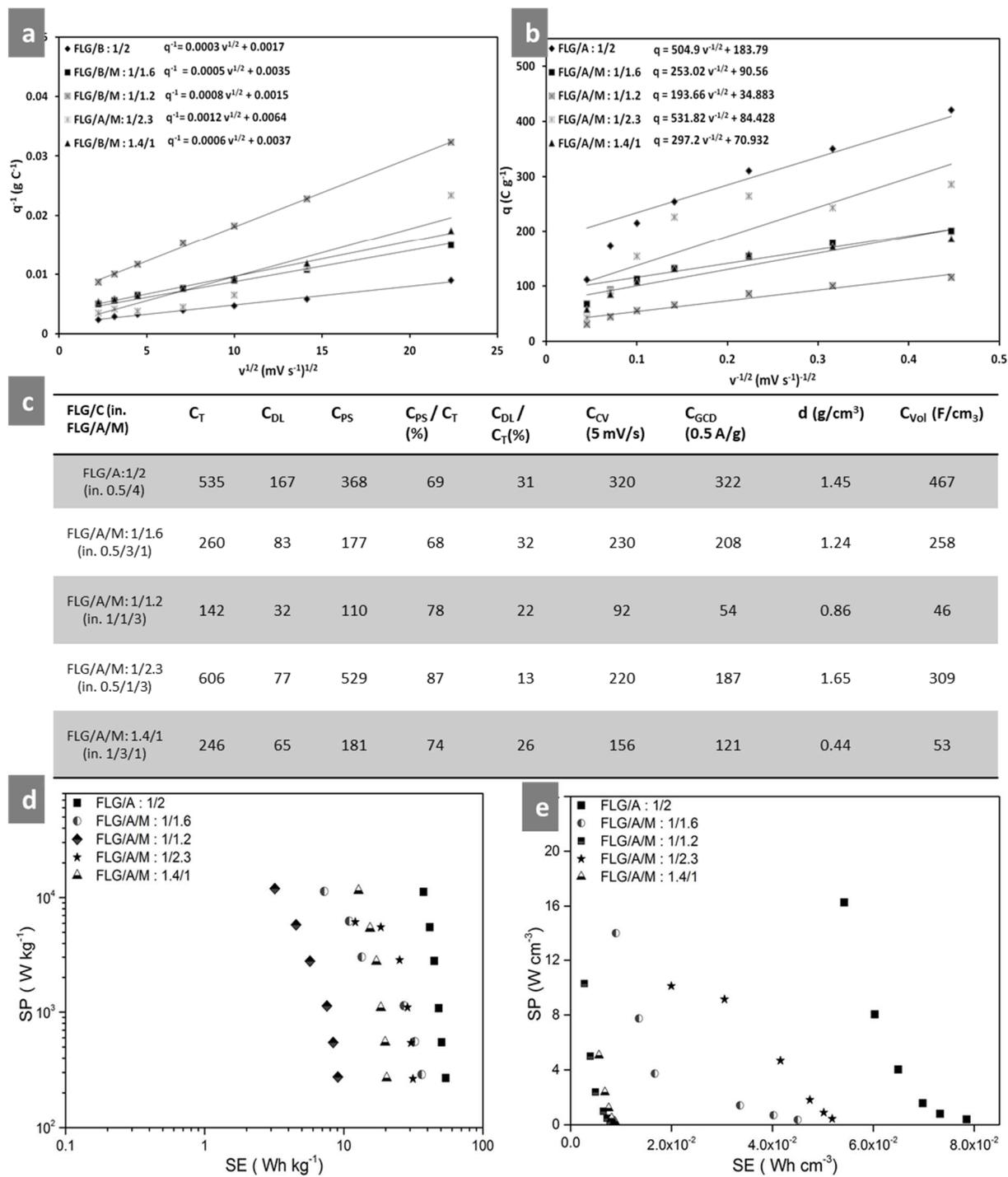

Figure Captions:

Figure 1. a) Exemplary TGA curves of two FLG/A/M composites after carbonization, b,c) the exemplary XPS C1s and N1s spectra of FLG/A/M: 1.4/1 before and after carbonization under $NH_3$ and Ar, d) table 2 containing the % of C, N, O elements according to XPS and/or elementary analysis*, e) deconvoluted N1s spectrum of FLG/A: 1/2 , f) % contribution of N-containing groups (blue-pyridinic; red-pyrolic; green-quaternary; purple-N-X) in different composites (*C1*-FLG/A:1/2; *C2*-FLG/A/M:1/1.6; *C3*-FLG/A/M:1/1.2; *C4*-FLG/A/M:1.4/1; *C5*-FLG/A/M:1.4/1-Ar).

Figure 2. Representative TEM and SEM micrographs of FLG/C composites: a) few microns size FLG flakes, b) edge of FLG flakes containing 8 sheets, c) FLG surface covered by C with some mesopores, d) FLG-C composite after carbonization, e) FLG/A/M composite (before carbonization) with visible hHLB spheres.

Figure 3. The representative SEM micrographs of FLG/A composite: a and b) before carbonization (side and top view), c) side view of FLG/C (FLG/A after carbonization, a.c.)), d and e) side-view of FLG/C (FLG/A a.c.) electrode, f) top-view of FLG/C (FLG/A a.c.) electrode.

Figure 4. a) Table 3 containing the total $S_{BET}$ and contribution of micropores $S_{BET}$, b) exemplary isotherms, c) pore size distribution curves, inset: mesoporous region, d) isotherm from $CO_2$ uptake with obtained $S_{BET}$ and $S_{micro}$, as well as DFT pore size distribution, for FLG/A:/1/2.

Figure 5. a) CV curves of FLG/A/(M) composites at 20 mV/s, b) GCD curves of FLG/A/(M) composites at 1A/g, c) a graph combining obtained specific gravimetric capacitance values, $C_{GCD}$ at different current densities, d) stability test for FLG/A composite at 20 A/g.

Figure 6. a-c) CV and d-i) GCD curves of FLG/A:1/2, FLG/A/M:1/1.6, FLG/A/M:1/2.3 at different potential sweeps.

Figure 7. a and b) Estimations of DL/Ps capacitance contribution: a) dependence of $q^{-1}$ vs. $v^{1/2}$, b) dependence of q vs. $v^{-1/2}$; c) table 4 regrouping the maximum total C from CV and DL, Ps contributions ($C_T$, $C_{DL}$, $C_{Ps}$, $C_{Ps}/C_T$, $C_{DL}/C_T$), capacitance from GCD ($C_{GCD}$), voltametric C ($C_{CV}$), packing density (d) and volumetric capacitance ($C_{vol}$); d and e) mass and volume normalized Ragone plots.

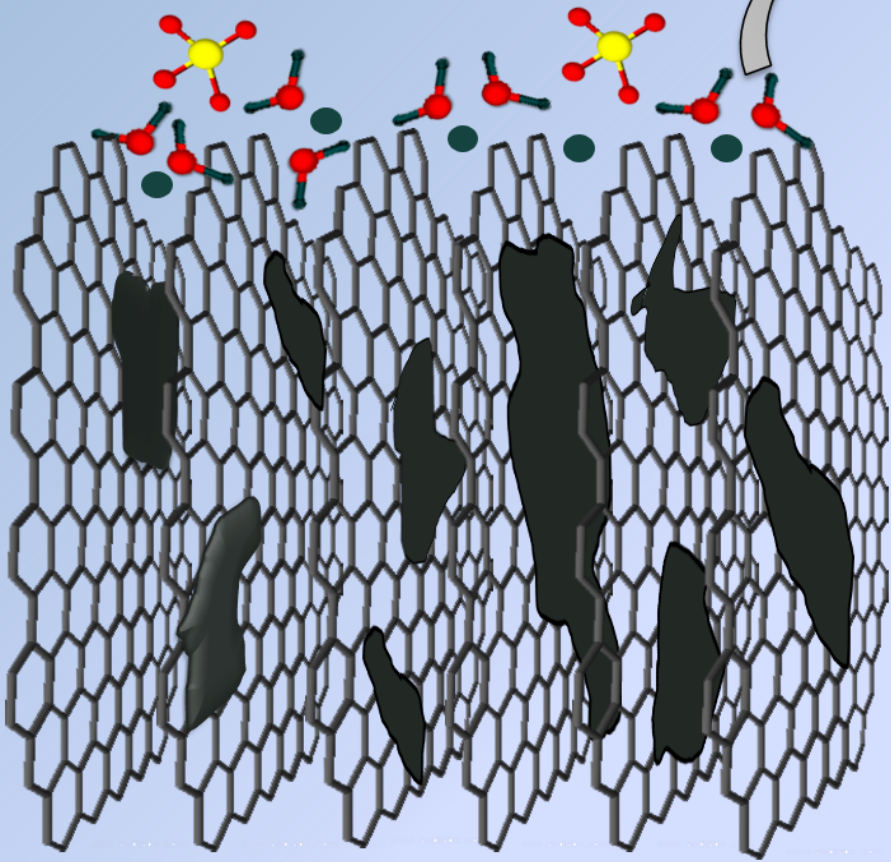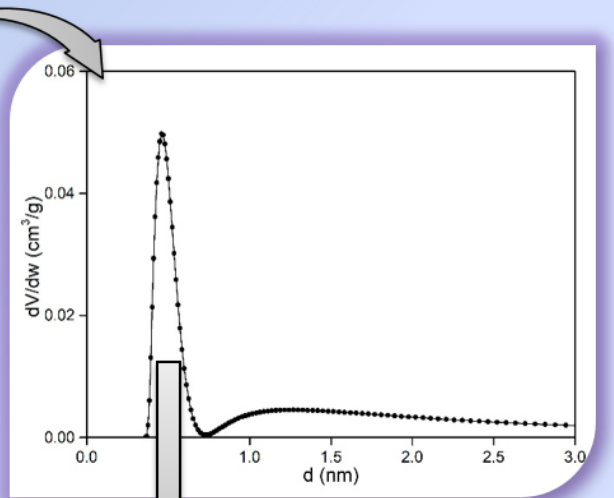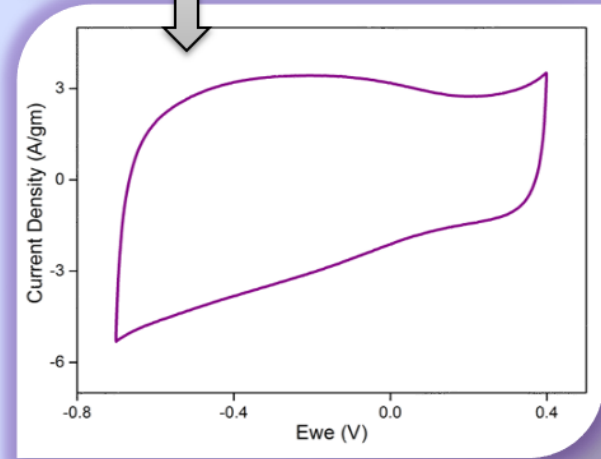